\documentstyle[11pt]{article}
\begin{document}

\title{Deformed Legendre Polynomial and Its Application \thanks{
This project supported by National Natural Science Foundation of China under
Grant 19771077 and LWTZ 1298.}}
\author{{Weimin Yang, Hu Li and Sicong Jing} \\
{\small Department of Modern Physics, University of Science and}\\
{\small Technology of China, Hefei, Anhui 230026, P.R.China}}
\maketitle

\begin{abstract}
A new kind of deformed calculus was introduced recently in studying of
parabosonic coordinate representation. Based on this deformed calculus, a
new deformation of Legendre polynomials is proposed in this paper, some
properties and applications of which are also discussed.
\end{abstract}

\bigskip \bigskip \bigskip

\section{Introduction}

Parastatistics was introduced by Green as an exotic possibility extending
the Bose and Fermi statistics \cite{s1} and for the long period of time the
interest to it was rather academic. Nowadays it finds some applications in
the physics of the quantum Hall effect \cite{s2} and (probably) it is
relevant to high temperature superconductivity \cite{s3}. The
paraquantization, carried out at the level of the algebra of creation and
annihilation operators, involves trilinear(or double) commutation relations
in place of the bilinear relations that characterize Bose and Fermi
statistics. Recently, the trilinear commutation relations of single
paraparticle systems was rewritten as bilinear commutation relations by
virtue of the so called R-deformed Heisenberg algebra \cite{s4}. For
instance, the trilinear commutation relations \cite{s5} 
\begin{equation}
\left[ a, \{ a^\dagger , a \} \right] = 2a, ~~~\left[ a, \{ a^\dagger,
a^\dagger \} \right] = 4a^\dagger, ~~~\left[ a, \{ a, a \} \right] = 0,
\end{equation}
where $a^\dagger$ and $a$ are parabose creation and annihilation operators
respectively, can be replaced by \cite{s6} 
\begin{equation}
[ a, a^\dagger ] = 1 + ( p-1 ) R, ~~~\{ R, a \} = \{ R, a^\dagger \} = 0,
~~~R^2 = 1,
\end{equation}
where $R$ is a reflection operator and $p$ is the paraquantization order $%
(p=1,2,3,...)$. Obviously, the bilinear commutation relations (2) may be
treated as some kind of deformation of the ordinary Bose commutator with
deformation parameter $p$.
\par
>From the experience of studying q-deformed oscillators \cite{s7}, we know
that it will be very useful if one introduces corresponding deformed
calculus to analyse this parabose system. This was done recently and based
on the new deformed calculus, the parabosonic coordinate representation was
developed \cite{s8}. Since special functions play important roles in
mathematical physics, it is reasonable to imagine that some deformation of
the ordinary special functions based on the new deformed calculus will also
play similar roles in studying the parabose systems. In this paper, we
introduce a new kind of deformation for the ordinary Legendre polynomials
and demonstrate its properties. As an example of applications of these
deformed Legendre polynomials, we discuss excitations on a parabose squeezed
vacuum state and calculate norm of the excitation states.
\par
The paper is organized as follows. In Section 2, for the sake of
self-contained of the present paper, we briefly mention the basic idea of
the new kind of deformed calculus. The deformed Legendre polynomials and
relevant differential expressions are introduced in section 3. Sections 4
and 5 are devoted to demonstration of orthonormality of the deformed
Legendre polynomials and their recursion relations respectively. In section
6, we show that the deformed Legendre polynomials can be used to normalize
the excitation states on a squeezed vacuum state for single parabose mode.
There are also some discussions and remarks in the last section.

\section{Deformed calculus related to parabosonic coordinate representation}

It is well-known that parabose algebra is characterized by the double
commutation relations (1). If one demands that the usual relations 
\begin{equation}
a = \frac{x+iP}{\sqrt{2}} ,~~~a^\dagger = \frac{x-iP}{\sqrt{2}}
\end{equation}
still work for the parabose case, where $x$ and $P$ stand for the coordinate
and momentum operator respectively, it can be proved that the most genaral
expression for the momentum operator $P$ in the coordinate $x$ diagonal
representation is of \cite{s5} \cite{s8} 
\begin{equation}
P = -i \frac{d}{dx} -i \frac{p-1}{2x} (1-R),
\end{equation}
where $p$ is the paraquantization order and $R$ the reflection operator
which has property $R f(x) = f(-x)$ in the coordinate representation for any 
$x$ dependent function $f(x)$. From (4) a new derivative operator $D$ can be
defined which acts on function $f(x)$ as 
\begin{eqnarray}
D f(x)&\equiv&\frac{D}{Dx} f(x) = \frac{d}{dx} f(x) + \frac{p-1}{2x} (1-R)
f(x)  \nonumber \\
&=&d\,f(x) + \frac{p-1}{2x} \left( f(x) - f(-x) \right) ,
\end{eqnarray}
where $d\,f=\frac{d}{d\,x}\,f$. Definition (5) implies that $D$ acts on an
even function $f_e (-x)= f_e (x)$ as the ordinary derivative $Df_e (x)= d\,
f_e (x)$, and $D$ acts on an odd function $f_o (-x)= -f_o (x)$ leads to $%
Df_o (x) = d\, f_o (x) + \frac{p-1}{x} f_o (x)$. For $p=1$ case, $D$ reduces
to the ordinary derivative operator $d$. Since $P= -iD$, Eq.(5) means that
the pair $(x,D)$ in realization of parabose algebra for a single degree of
freedom plays the same role as $(x,d)$ in realization of the ordinary Bose
algebra. Like q-deformed calculus in which the q-analogue of the number
system was defined by $[n]_q = \frac{q^n -1}{q-1}$ \cite{s7}, such that when 
$q \rightarrow 1$, $[n]_q \rightarrow n$, in the present case, one can
introduce a new kind of deformed number system which is defined by 
\begin{equation}
[n] = n + \frac{p-1}{2} (1 - (-)^n ).
\end{equation}
Obviously, $[2k] = 2k$, $[2k+1] = 2k+p$ for any integer $k$ and when $p
\rightarrow 1$, $[n] \rightarrow n$. So paraquantization order $p$ may be
referred to as a deformation parameter. In terms of the number system $[n]$,
basis vectors of Fock space for single mode of parabose oscillators take the
usual form 
\begin{equation}
|n \rangle = \frac{(a^\dagger)^n}{\sqrt{[n]!}} |0 \rangle, ~~~ a^\dagger |n
\rangle = \sqrt{[n+1]} |n+1 \rangle, ~~~ a |n \rangle = \sqrt{[n]} |n-1
\rangle,
\end{equation}
where $[n]!=[n][n-1]...[1], [0]! \equiv 1$, and $|0 \rangle$ is the unique
vacuum vector satisfying $a |0 \rangle =0, aa^\dagger |0 \rangle = p |0
\rangle$. Generalization of the ordinary differential relation $d x^n =
nx^{n-1} $ reads 
\begin{equation}
D x^n = [n] x^{n-1},
\end{equation}
which reveals the effect of the deformed derivative operator $D$ on the
polynomials of $x$. It is worthy of metion that in some special situation
the usual Leibnitz rule also works for the deformed operator $D$ 
\begin{equation}
D(f\,g) = (Df)\,g + f\,(Dg),
\end{equation}
where either $f(x)$ or $g(x)$ is an even function of $x$.
\par
Of course, inversion of the deformed derivative operator $D$ also leads to a
new deformed integration which may be formally written as \cite{s8} 
\begin{eqnarray}
\!& &\! \int \,Dx F(x) = \sum_{n=0}^{\infty} (-)^n \left( \int \,dx \frac{p-1%
}{2x} (1-R) \right)^n \int \,dx F(x)  \nonumber \\
\!&=&\! \int \,dx F(x) - \int \,dx \frac{p-1}{2x} (1-R) \int \,dx F(x) 
\nonumber \\
\,& &\! +\left( \int \,dx \frac{p-1}{2x} (1-R) \right)^2 \int \,dx F(x) -
\cdots .
\end{eqnarray}
>From this expression, it is easily seen that if $F(x)$ is an odd function of 
$x$, its deformed integration will reduce to the ordinary integration, that
is, $\int \,Dx F(x) = \int \,dx F(x)$ for $F(-x)=-F(x)$. Corresponding to
Eq. (8), one has 
\begin{equation}
\int \,Dx \,x^n = \frac{x^{n+1}}{[n+1]} + c,
\end{equation}
where c is an integration constant. Eq.(10) gives a formal definition for the
deformed integration in the sence of indefinite integral. For definite
integral, we have 
\begin{eqnarray}
\!& &\! \int_a^b \,Dx F(x) = \int_a^b \,dx \sum_{n=0}^{\infty} (-)^n \left( 
\frac{p-1}{2x} (1-R) \int_a^x \,dx \right)^n F(x)  \nonumber \\
\!&=&\! \int_a^b \,dx F(x) - \int_a^b \,dx \frac{p-1}{2x} (1-R) \int_a^x
\,dx F(x)  \nonumber \\
\!& &\! + \int_a^b \,dx \frac{p-1}{2x} (1-R) \int_a^x \,dx \frac{p-1}{2x}
(1-R) \int_a^x \,dx F(x) - \cdots .
\end{eqnarray}
If either $F(x)$ or $G(x)$ is an even function of $x$, one has a formula of
integration by parts from Eq.(9) 
\begin{equation}
\int_a^b \,Dx \frac{D\,F}{Dx} \,G = FG |_a^b - \int_a^b \,Dx F\,\frac{D\,G}{%
Dx}.
\end{equation}

\section{Defprmed Legendre polynomials and their differential expressions}

Let us consider solutions of deformed Legendre equation based on the
deformed derivative operator $D$ defined in the previous section 
\begin{equation}
(1-x^2)D^2f(x)-2xDf(x)+\mu f(x)=0,
\end{equation}
or according to Eq.(9), the deformed Legendre equation is equivalent to 
\begin{equation}
D\left( (1-x^2)Df(x)\right) +\mu f(x)=0.
\end{equation}
In terms of the ordinary derivative notation, Eq.(14) can be rewriiten as 
\[
(1-x^2)\frac{d^2}{dx^2}f(x)-\left( 2x-(p-1)(\frac 1x-x)\right) \frac
d{dx}f(x)-\frac{p-1}2\left( 1+\frac 1{x^2}\right) f(x)
\]
\begin{equation}
+\frac{p-1}2\left( 1+\frac 1{x^2}\right) f(-x)=-\mu f(x),
\end{equation}
which will reduce to the usual Legendre equation when $p\rightarrow 1$. We
find out that when the parameter $\mu $ takes eigenvalues $\mu
=[n][n+1],n=0,1,2,3,...$, for each given paraquantization order $p$, the
deformed Legendre equation has bounded solutions (eigenfunctions) within a
whole closed interval $-1\leq x\leq 1$ which form a set of orthonormal
functions in the interval. In fact, it is not difficult to see that the
following polynomials 
\begin{equation}
P_n(x)=\sum_{k=0}^{[n/2]^{^{\prime }}}\frac{(-)^k[2n-2k]!}{2^nk!(n-k)![n-2k]!%
}x^{n-2k},
\end{equation}
where $[k]^{^{\prime }}$ in the above of summation notation $\sum $ stands
for the largest integer smaller than or equal to $k$, are the desired
solutions of the deformed Legendre equation for $\mu =[n][n+1]$ which will
reduce to the usual Legendre polynomials when $p\rightarrow 1$. One can
substitute Eq.(17) into the deformed Legendre equation and check
coefficients of all powers of $x$ being zero. So the polynomials (17) may be
considered as a deformation of the usual Legendre polynomials. The first few
polynomials of $P_n(x)$ have the following explicit forms 
\begin{eqnarray}
&&P_0(x)=1,~~~P_1(x)=x,~~~P_2(x)=\frac 12([3]x^2-[1]),  \nonumber \\
&&P_3(x)=\frac 12([5]x^3-[3]x),~~~P_4(x)=\frac
18([5][7]x^4-2[3][5]x^2+[1][3]),  \nonumber \\
&&P_5(x)=\frac 18([7][9]x^5-2[5][7]x^3+[3][5]x),  \nonumber \\
&&P_6(x)=\frac 1{48}([7][9][11]x^6-3[5][7][9]x^4+3[3][5][7]x^2-[1][3][5]).
\end{eqnarray}
Also from Eq.(17) we know that $P_n(-x)=(-)^nP_n(x)$, which means that the
deformed Legendre polynomial $P_n(x)$ has its parity $(-)^n$.
\par
We would like to point out a differential expression for the deformed
Legendre polynomials $P_n (x)$ to conclude this short section 
\begin{equation}
P_n (x) = \frac{1}{2^n n!} D^n (x^2-1)^n.
\end{equation}
which is similar to Rodrigues formula for the ordinary Legendre polynomials
and can be proved straightforwardly.

\section{Orthonormality of $P_n (x)$}

Firstly we show the orthogonality of the deformed Legendre polynomials 
\begin{equation}
\int_{-1}^{1} \,Dx P_n (x) P_m (x) =0, ~~~(n \neq m).
\end{equation}
For $n+m$ odd case, Eq.(20) works obviously. In fact, from the parity of $%
P_n (x)$ we have $P_n (-x) P_m (-x) =(-)^{n+m} P_n (x) P_m (x) = -P_n (x)
P_m (x) $, which means that the deformed integration in Eq.(20) will reduce
to an ordinary integration with an odd integrand $P_n (x) P_m (x)$ over an
interval of integration $[-1,1]$, therefore the integration should be zero.
For $n+m$ even case, $P_n (x)$ and $P_m (x)$ satisfy the following equations 
\begin{equation}
D \left( (1-x^2) DP_n (x) \right) + \mu_n P_n (x) = 0,
\end{equation}
\begin{equation}
D \left( (1-x^2) DP_m (x) \right) + \mu_m P_m (x) = 0,
\end{equation}
respectively, where $\mu_n = [n][n+1], \mu_m =[m][m+1]$. Multiplying Eq.(21)
and Eq.(22) by $P_m (x)$ and $P_n (x)$ respectively, and substracting the
resulting equations, then integrating it over $[-1,1]$, we have 
\[
\int_{-1}^{1} \,Dx \left( P_m \frac{D}{Dx} \left( (1-x^2) \frac{D}{Dx} P_n
\right) - P_n \frac{D}{Dx} \left( (1-x^2) \frac{D}{Dx} P_m \right) \right) 
\]
\begin{equation}
+ (\mu_n -\mu_m ) \int_{-1}^{1} \,Dx P_n P_m = 0.
\end{equation}
Since the first integration in Eq.(23) satisfies condition of the deformed
integration by parts either for $n$ and $m$ are all even or all odd, we can
write it as 
\begin{eqnarray}
& &\int_{-1}^{1} \,Dx \frac{D}{Dx} \left( (1-x^2) (P_m \frac{D}{Dx} P_n -
P_n \frac{D}{Dx} P_m ) \right)  \nonumber \\
&=&(1-x^2) (P_m \frac{D}{Dx} P_n - P_n \frac{D}{Dx} P_m ) |_{-1}^{1} = 0,
\end{eqnarray}
here we have used a fact that $P_n (x)$, $P_m (x)$ and their deformed
derivatives are all polynomials of $x$ and are bounded at $x= \pm 1$ for a
fixed paraquantization order $p$. Noticing that $\mu_n \neq \mu_m$, so
Eq.(20) is proved.
\par
Then we calculute integration 
\begin{equation}
N_n^2 = \int_{-1}^{1} \,Dx P_n (x) P_m (x).
\end{equation}
Substituting Eq.(19) into this integration, we have 
\begin{equation}
N_n^2 = \frac{1}{2^n n!} \int_{-1}^{1} \,Dx P_n (x) \frac{D^n}{Dx^n} (x^2
-1)^n.
\end{equation}
Noticing that the integration (26) also satisfies the condition of the
deformed integration by parts no matter what non-negative integer $n$ is, we
get 
\[
N_n^2 = \frac{1}{2^n n!} \left( P_n (x) \frac{D^{n-1}}{Dx^{n-1}} (x^2 -1)^n
|_{-1}^{1} - \int_{-1}^{1} \,Dx \frac{D \,P_n (x)}{Dx} \frac{D^{n-1}}{%
Dx^{n-1}} (x^2 -1)^n \right). 
\]
It is not difficult to see that for $m \leq n$, $\frac{D^m}{Dx^m} (x^2 -1)^n
= 0$ at $x=\pm 1$, so 
\[
N_n^2 = \frac{-1}{2^n n!} \int_{-1}^{1} \,Dx \frac{D \,P_n (x)}{Dx} \frac{
D^{n-1}}{Dx^{n-1}} (x^2 -1)^n. 
\]
Continuing the procedure of integration by parts, at last we arrive at 
\begin{equation}
N_n^2 = \frac{(-)^n}{2^n n!} \int_{-1}^{1} \,Dx \frac{D^n \,P_n (x)}{Dx^n}
(x^2 -1)^n = \frac{(-)^n [2n]!}{2^n n! 2^n n!} \int_{-1}^{1} \,Dx (x^2 -1)^n.
\end{equation}
Using Eq.(11) we can rewrite the integral in the right-hand side of Eq.(27)
as 
\[
\int_{-1}^{1} \,Dx (x^2 -1)^n = 2 \sum_{k=0}^{n} (-)^{n-k} C_n^k \frac{1}{
[2k+1]} = 2(-)^n \sum_{k=0}^{n} (-)^k C_n^k \frac{1}{2k+p} 
\]
\[
= 2(-)^n \sum_{k=0}^{n} C_n^k \int_0^1 \,dx x^{2k+p-1} = 2(-)^n \int_0^1
\,dx x^{p-1} (1-x^2)^n = \frac{(-)^n 2^{2n+1} (n!)^2}{[2n+1]!}, 
\]
where $C_n^k = \frac{n!}{k!(n-k)!}$. Substituting this result into Eq.(27)
we obtain 
\begin{equation}
N_n^2 = \frac{2}{[2n+1]} ,
\end{equation}
which is called the norm of $P_n (x)$. Combining Eqs.(20)and (28) we have 
\begin{equation}
\int_{-1}^{1} \,Dx P_n (x) P_m (x) = \frac{2}{[2n+1]} \delta_{n,m} .
\end{equation}
Thus we demonstrated the orthonormality of the deformed Legendre polynomials 
$P_n (x)$.

\section{Recursion relations of $P_n (x)$}

There are some definite relations between neighbouring deformed Legendre
polynomials and their derivatives which are called recursion relations of $%
P_n(x)$. The main recursion relations are the following three: 
\begin{equation}
\lbrack n+1]P_{n+1}(x)-[2n+1]xP_n(x)+[n]P_{n-1}(x)=0,
\end{equation}
\begin{equation}
DP_{n+1}(x)-xDP_n(x)-[n+1]P_n(x)=0,
\end{equation}
\begin{equation}
xDP_n(x)-DP_{n-1}(x)-[n]P_n(x)=0.
\end{equation}
It is straightforward to prove these relations by virtue of the definition
(17) and the deformed differential relation (8). For instance, one can
substitute (17) into (30) and obtain a polynomial of $x$, then calculate
coefficient for each power of $x$ and find that all of these coefficients
will be zero, thus relation (30) is true. Similarly, one can convince
oneself that relations (31) and (32) also work. Based on the above three
relations, one has more recursion relations for the deformed Legendre
polynomials 
\begin{equation}
DP_{n+1}(x)-DP_{n-1}(x)-[2n+1]P_n(x)=0,
\end{equation}
\begin{equation}
(x^2-1)DP_n(x)-[n]xP_n(x)+[n]P_{n-1}(x)=0.
\end{equation}
Obviously, these recursion relations will reduce to ones of the usual
Legendre polynomials when $p\rightarrow 1$.
\par
The above recursion relations are very useful for calculating deformed
integrations with the deformed Legendre polynomials such as 
\[
\int_{-1}^{1} \,Dx \,xP_m (x) P_n (x). 
\]
>From Eq.(30), one has 
\[
xP_m (x) = \frac{[m+1]}{[2m+1]} P_{m+1} (x) + \frac{[m]}{[2m+1]} P_{m-1}
(x), 
\]
so that 
\[
\int_{-1}^{1} \,Dx \,xP_m (x) P_n (x) = \frac{[m+1]}{[2m+1]} \int_{-1}^{1}
\,Dx P_{m+1} (x) P_n (x) 
\]
\[
+ \frac{[m]}{[2m+1]} \int_{-1}^{1} \,Dx P_{m-1} (x) P_n (x). 
\]
Furthermore, by virtue of the orhtonormality relation (29), one gets 
\[
\int_{-1}^{1} \,Dx \,x P_m (x) P_n (x) = \left\{ 
\begin{array}{cc}
\frac{2[n]}{[2n-1][2n+1]} , & m=n-1, \\ 
\frac{2[n+1]}{[2n+1][2n+3]} , & m=n+1, \\ 
0, & m-n \neq \pm 1.
\end{array}
\right. 
\]

\section{Some application and discussion}

As an application of the deformed Legendre polynomials $P_n (x)$, we want to
point out that there exist some states in parabose radiation field whose
normalizing factors are related to $P_n (x)$. To see this, let us consider
excitations on a parabose squeezed vacuum state for a single mode case.
Denoting $|r,0 \rangle = S(r) |0 \rangle$ as the parabose squeezed vacuum
state ($r$ is a real number), where
$S (r) = (sech\, r)^{p/2} exp \left( \frac{1}{2}
(a^\dagger)^2 tanh\, r \right)$, we introduce 
\begin{equation}
|r,m \rangle = (a^\dagger)^m |r,0 \rangle ~~~(m=1,2,3,...)
\end{equation}
as such kind of excitation states. It is easily to see that the parabose
squeezed vacuum state $|r,0 \rangle$ is normalized $\left( \langle r,0|r,0
\rangle = 1 \right)$ and the parabose squeezed excitation states have not
been normalized. In order to normalize these states, we would like to prove
by induction 
\begin{equation}
\langle r,m|r,m \rangle = [m]! (cosh\, r)^m P_m (cosh\, r) ,
\end{equation}
where $P_m (cosh\, r)$ is the deformed Legendre polynomial with argument $
x=cosh\, r$. Firstly, we have 
\begin{eqnarray}
a |r,0 \rangle &=& (sech\, r)^{p/2} a e^{\frac{(a^\dagger)^2}{2} tanh \,r} |0
\rangle    \nonumber\\
&=& (sech\, r)^{p/2} e^{\frac{(a^\dagger)^2}{2} tanh\, r}
(a + a^\dagger tanh\, r) |0 \rangle 
= a^\dagger tanh\, r |r,0 \rangle .
\end{eqnarray}
Using the R-deformed commutation relation (2) for paraboson and noticing
that $R |0 \rangle = |0 \rangle$, we find 
\[
\langle r,1|r,1 \rangle = \langle r,0|a a^\dagger|r,0 \rangle = \langle r,0|
\left( 1 + a^\dagger a + (p-1)R \right) |r,0 \rangle 
\]
\begin{equation}
= p + tanh^{2}{r} \langle r,1|r,1 \rangle = [1]! cosh\, r P_1 (cosh\, r).
\end{equation}
Then supposing Eq.(36) is true for $n \leq m$, that is, 
\begin{equation}
\langle r,n-1|r,n-1 \rangle = [n-1]! (cosh\, r)^{n-1} P_{n-1} (cosh\, r) ,
\end{equation}
we show that Eq.(36) works. In fact, using the following relations 
\begin{eqnarray}
\left[a, (a^\dagger)^n\right]&=& (a^\dagger)^{n-1}
\left(n+\frac{p-1}{2}(1-(-)^n)R \right), \nonumber\\
\left[a^\dagger, a^n\right]&=& -a^{n-1} \left(n+\frac{p-1}{2}(1-(-)^n)R
\right),
\end{eqnarray}
we have 
\begin{eqnarray*}
& &\langle r,m|r,m \rangle = \langle r,0|a^{m-1}aa^\dagger (a^\dagger)^{m-1}
|r,0 \rangle \\
& &= \langle r,0|a^{m-1} \left( 1+a^\dagger a + (p-1)R \right)
(a^\dagger)^{m-1} |r,0 \rangle \\
& &= \langle r,m-1|r,m-1 \rangle +(p-1)(-)^{m-1} \langle r,m-1|r,m-1 \rangle
\\
& &\,\,+ \langle r,0| \left( a^\dagger a^{m-1} + [m-1]a^{m-2} \right) \left(
(a^\dagger)^{m-1} a + [m-1](a^\dagger)^{m-2} \right) |r,0 \rangle \\
& &=tanh^{2}{r} \langle r,m|r,m \rangle -[m-1]^2 \langle r,m-2|r,m-2 \rangle
\\
& &\,\,+ \left(2[m-1] +1 +(p-1)(-)^{m-1} \right) \langle r,m-1|r,m-1 \rangle ,
\end{eqnarray*}
or 
\begin{eqnarray}
& &\langle r,m|r,m \rangle = - cosh^{2}{r} [m-1]^2 \langle r,m-2|r,m-2
\rangle  \nonumber \\
& & + cosh^{2}{r} ( 2[m-1] +1 + (p-1)(-)^{m-1} ) \langle r,m-1|r,m-1 \rangle
..
\end{eqnarray}
Substituting (39) into (41), we get 
\begin{eqnarray*}
& &\langle r,m|r,m \rangle = - cosh^{m}{r} [m-1]! [m-1]P_{m-2} (cosh\, r) 
\nonumber \\
& &\,\,+cosh^{m+1}{r}[m-1]!(2[m-1] +1 +(p-1)(-)^{m-1} ) P_{m-1} (cosh\, r) .
\end{eqnarray*}
Noticing that $2[m-1] +1 + (p-1)(-)^{m-1} =[2m-1]$ and using the recursion
relation (30) for $P_n (x)$, we finally arrive at (36). Thus we see that the
deformed Legendre polynomials indeed can be used to normalize the excitation
states on a squeezed vacuum state for a single parabose mode.
\par
In summary, we introduced a new kind of deformation for the usual Legendre
polynomials and discussed their main preperties in this paper. These
deformed Legendre polynomials may have some applications in studying
parabose systems. For instance, they can be used to describe excitations on
a parabose squeezed vacuum state. Comparing with the case of ordinary
Legendre polynomials, one will naturally ask a question: is there any
genarating function of the deformed Legendre polynomials and what it is?
Work for answering these questions is on progress.


\begin{thebibliography}{9}
\bibitem{s1}  H. S. Green, {\em Phys.Rev.} {\bf 90},270 (1953).

\bibitem{s2}  B. I. Halperin, {\em Phys.Rev.Lett.} {\bf 52} 1583 (1984).

\bibitem{s3}  F. Wilczek, {\em Fractional Statistics and Anyon 
Superconductivity} (World Scientific, Singapore, 1990).

\bibitem{s4}  M. A. Vasiliev, {\em Int.J.Mod.Phys.} {\bf A6} 1115 (1991); 
T. Brzezinski, I. L. Egusquiza and A. J. Macfarlane, {\em Phys.  Lett.} {\bf %
B 311} 202 (1993);  M. S. Plyushchay, {\em Nucl.Phys.} {\bf B491} 619 (1997).

\bibitem{s5}  Y. Ohnuki and S. Kamefuchi, {\em Quantum Field Theory and 
Parastatistics} (Springer-Verlag, 1982).

\bibitem{s6}  A. J. Macfarlane, {\em J.Math.Phys.} {\bf 35} 1054 (1994).

\bibitem{s7}  L. C. Biedenharn, {\em J.Phys.A:Math.Gen.} {\bf 22} L873
(1989);  A. J. Macfarlane, {\em J.Phys.A:Math.Gen.} {\bf 22} 4581 (1989); 
C. Sun and H. Fu, {\em J.Phys.A:Math.Gen.} {\bf 22} L983 (1989);  M. Arik
and D. Coon, {\em J.Math. Phys.} {\bf 17} 524 (1976).

\bibitem{s8}  S. Jing, {\em J. Phys.A:Math.Gen.} {\bf 31} 6347 (1998).
\end{thebibliography}
\end{document}